\documentclass[10pt]{article}
\usepackage{amssymb}
\usepackage{amstext}
\newcommand*\de{\mathrm{d}}

\renewcommand*\epsilon{\varepsilon}
\renewcommand*\phi{\varphi}
\renewcommand*\theta{\vartheta}
\begin{document}

\title{\bf \large Modified Ostrogradski formulation of  field theory }

\author{M. Leclerc \\ \small Section of Astrophysics and Astronomy,
Department of Physics, \\ \small University of Athens, Greece}
\date{\small October 2, 2006}
\maketitle
 
\begin{abstract}
We present a method for the Hamiltonian formulation of field theories
that are based on Lagrangians containing second derivatives. 
The new feature of 
our formalism is that all four partial derivatives of the field 
variables are initially considered as independent fields, in contrast to  
the conventional Ostrogradski method, 
where only the velocity is turned into an independent 
field variable. 
The consistency of the formalism is demonstrated by 
simple  unconstrained and constrained second order scalar field 
theories. Its application to General Relativity is briefly outlined. 
\end{abstract}
 
\maketitle

\section{Introduction}

There are two main properties one usually requires a Hamiltonian to possess. 
First, it should be a conserved quantity, the energy, and second, it should 
generate the time evolution of the system under consideration. Both 
features are intimately related, since in relativistic theories, the energy 
can be viewed as the variable  canonically conjugate to time. However, since 
the Hamiltonian formalism is based on a 3+1 split of spacetime, this relation 
is not always directly obvious. As a matter of fact, in classical mechanics, 
but also in field theory, 
the Hamiltonian is usually 
introduced via  a Legendre transformation of the Lagrangian, and the relation 
to the energy (i.e., to the time component of the integrated stress-energy 
tensor, in the field theory case) is only  established afterwards and appears 
rather as a coincidence. 

Here instead, 
we choose to proceed the other way around, that is, we start from 
the expression for the energy and try to find a set of canonical 
phase space variables such that the Hamiltonian that emerges after 
substitution of those variables into the initial expression, does indeed 
generate the time evolution of the system. 

The stress-energy tensor for second order theories 
can be derived with  Noether's theorem 
(see our review paper \cite{leclerc}) from invariance of the 
theory under coordinate transformations. Assume that the 
fields, collectively denoted with $\phi$,  transform as 
\begin{equation} \label{1}
\delta \phi = \xi^i \phi_{,i} + \frac{1}{2} \xi^i_{\ ,k} 
(\sigma \phi)^k_{\  i}, 
\end{equation}  
where  $(\sigma \phi)^k_{\ i}$ denotes the action of  
the generators of the general linear 
group on $\phi$. The explicit form depends 
on the scalar, vector or tensor nature of the fields $\phi$. 
(In other words, $\delta \phi$ is 
 the Lie derivative of the field with respect to $\xi^i$.) 
Invariance under global transformations $\xi^i = const$ leads to 
$\tau^k_{\ i,k} = 0 $, where the canonical 
stress-energy tensor is defined as   
\begin{equation} \label{2} 
\tau^k_{\ i} = \left[\frac{\partial \mathcal L}{\partial \phi_{,k}} 
- (\frac{\partial \mathcal L}{\partial
  \phi_{,k,l}})_{,l}\right]  \phi_{,i} 
+ 
\frac{\partial \mathcal L}{\partial \phi_{,k,l}} \phi_{,i,l} 
- \delta^k_i \mathcal L.  
\end{equation}
It can further be shown (see \cite{goldberg} or  \cite{leclerc}) 
that for generally covariant  
Lagrangians, $\tau^k_{\ i}$ can be brought into the form 
\begin{equation} \label{3} 
\tau^k_{\ i} = 
\left[-\frac{\partial \mathcal L}{\partial \phi_{,k,l}} \phi_{,i} 
- \frac{1}{2}
\frac{\partial \mathcal L}{\partial \phi_{,l}} (\sigma \phi)^k_{\ i}
-  \frac{\partial \mathcal L}{\partial \phi_{,m,l}}
[(\sigma \phi)^k_{\ i}]_{,m}\right]_{,l} 
 + \frac{1}{2} \left[\frac{\partial \mathcal L}{\partial \phi_{,m,l}} 
(\sigma \phi)^k_{\ i}\right]_{,m,l},  
\end{equation}
where the expression in the first bracket was shown to be 
antisymmetric in $kl$, 
and the totally symmetric part (in $mlk$) of the expression in the second 
bracket was shown to be  zero. 
As a result, $\tau^k_{\ i,k} = 0$ identically, in 
accordance with  the second Noether theorem. 

In first order theories, it is convenient to introduce four 
momentum variables $\pi^{(i)}  = \partial \mathcal L/ \partial \phi_{,i}$, 
and then perform the 3+1 split with the help of a timelike 
unit vector $n^i$, which specifies the direction of the time evolution. 
The physical momentum is then given by $\pi = \pi^{(i)} n_i$ (and the 
velocities by $\phi_{,i} n^i$). In this way, one can set up an 
explicitely  covariant Hamiltonian formulation of the theory (see 
\cite{leclerc2}). Throughout this paper, we assume $n_i = \delta_i^0$, 
such that $\pi = \pi^{(0)}$. The introduction of 
$\pi^{(i)}$ is nevertheless useful, since it allows to write 
the stress-energy tensor in the simple form $\pi^{(k)}\phi_{,i} - \delta^k_i 
\mathcal L$. 

Based on those considerations, it is natural, in the framework 
of second order theories, to introduce the following momenta 
\begin{equation} \label{4}
\pi^{(i)} = \frac{\partial \mathcal L}{\partial \phi_{,i}} - (
\frac{\partial \mathcal L}{\partial \phi_{,k,i}})_{,k}, \ \ \ \  
p^{m(i)}= \frac{\partial \mathcal L}{\partial \phi_{,m,i}},  
\end{equation}
which contain the physical momenta (i.e., along $n_i = \delta_i^0$) 
\begin{equation}   \label{5} 
\pi = \pi^{(0)} = 
\frac{\partial \mathcal L}{\partial \dot \phi} - (
\frac{\partial \mathcal L}{\partial \dot \phi_{,k}})_{,k}, \ \ \ \  
p^{m}= p^{m(0)} = \frac{\partial \mathcal L}{\partial \dot \phi_{,m}},  
\end{equation} 
where the dot denotes partial time derivation. Note that we use 
latin letters for spacetime indices and greek letters for spatial indices
(and zero for the time component). 
Further, we  take the convention 
that expressions of the form $\partial \mathcal  L/ \partial \phi_{,i,0}$ 
or equivalently, $\partial \mathcal L/ \partial \dot \phi_{,i}$,  
are always to be interpreted as the  component $m = 0$ 
of the initial expression $\partial \mathcal L/ \partial \phi_{,i,m}$.  
(This is very important, for instance if 
$\mathcal L = \phi_{,i,m} \phi^{,i,m}$. With our convention, we have  
$\partial \mathcal L/ \partial \phi_{,\mu,0} = 2 \phi^{,\mu,0}$, although 
the mixed term  $\phi_{,\mu,0}\phi^{,\mu,0}$ is  actually contained twice in 
$\mathcal L$.)   

The stress-energy tensor (\ref{2}) can now be written in the form 
\begin{equation} \label{6}
\tau^k_{\ i} = \pi^{(k)} \phi_{,i} + p^{m(k)} \psi_{m,i} - 
\delta^k_i \mathcal L, 
\end{equation}
where it is tempting to see in $\psi_m \equiv \phi_{,m}$ 
the variable canonically 
conjugate to $p^{m(i)}$. Further, the identically conserved form of the 
stress-energy tensor from (\ref{3}) takes the  form 
\begin{equation} \label{7}
\tau^k_{\ i} = - \frac{1}{2}\left[\pi^{(m)} (\sigma \phi)^k_{\ i} + 
p^{j(m)} (\sigma \psi_j)^k_{\ i}\right]_{,m}
 \end{equation}
where $(\sigma \psi_j)^k_{\ i} 
= 2 \delta^k_j \phi_{,i} + [(\sigma \phi)^k_{\ i}]_{,j}$, 
i.e., it acts correctly on $\psi_j$ in accordance with its total 
tensor structure, taking account of the additional vector index.  

The last relation holds only in generally covariant theories (e.g., 
General Relativity), and is given here only to demonstrate how naturally 
one is led to the specific choice of canonical variables.   
For the 
moment, we confine ourselves to special relativistic 
theories. 

The conserved momentum is given as integral over a three dimensional 
 hypersurface,  
$\mathcal P_i = \int \tau^k_{\ i} \de \sigma_k$. For $n_i = \delta_i^0$, 
this is simply $\mathcal P_i = \int \tau^0_ {\ i} \de^3 x$. We see that 
only the physical momenta   (\ref{5}) enter that expression. 
In particular, for the field energy $\mathcal H = \mathcal P_0$, 
we find 
\begin{equation} \label{8} 
\mathcal H = \int \left[ \pi \dot \phi + p^m \dot \psi_{m} - \mathcal
  L\right] \de^3 x, 
\end{equation} 
 where we use already the letter $\mathcal H$ although we yet  have to 
eliminate the velocities. 

Until this point, we have done nothing but written down the 
canonical field energy in terms of certain variables. Whether or not 
we can indeed interpret $\phi, \pi$ and $\psi_m, \pi^m$ as canonical 
 phase space variables and whether or not $\mathcal H$ 
generates the time evolution 
of those variables is still an open question. However, from the above
relations, any different choice of variables would seem unnatural.

The conventional Hamiltonian formulation of second order theories is 
based on two pairs of canonically conjugated 
variables,  $\phi, \pi$ and $\psi, \pi$, with $\psi= \dot \phi $. 
This method goes back to Ostrogradski. We will investigate the 
relation between both methods at a later stage of this paper. Let us 
just remark that the fact that we use more variables should ultimately result 
in a theory with more constraints. If this is not the case, then both methods 
cannot be equivalent. 

Finally, we assume the following equal-time Poisson brackets  
\begin{equation} \label{9} 
[\phi(x), \pi(y)] = \delta (x-y), \ \ 
[\psi_m(x), p^{k}(y)] = \delta_m^k \delta (x-y),
\end{equation}
and zero for any other bracket, e.g., $[\phi, \psi_m] = 0$. 
For simplicity, we use $x,y$ to denote the space points $x^{\mu},
y^{\mu}$, and $\delta (x-y) = \delta^{(3)}(x-y)$ for  the three dimensional 
delta function. The covariant form (i.e., with respect to a general, 
unspecified hypersurface) of those relations are given in \cite{leclerc}. 
Note that even classically, the relations (\ref{9}) cannot be 
\textit{verified}. It is indeed possible to construct the Poisson bracket  
such that it gives canonical relations for an arbitrary choice 
of variables. The only thing subject to verification is the resulting 
Hamiltonian theory. To this we turn now. 

In the next two sections, we will apply the above formalism to several simple 
second order scalar field Lagrangians. It turns out that in all cases, 
the formalism reveals itself to be consistent. In section \ref{ostro}, 
we briefly analyze the relation to the conventional Ostrogradski formalism. 
Finally, in section \ref{gr}, we treat General Relativity as a 
constrained second order theory and in section \ref{diss}, we 
briefly discuss the relation 
of the first class constraints to the diffeomorphism invariance of the 
theory.

\section{Unconstrained theories} \label{unconstr}

As a first example, we consider the Lagrangian 
\begin{equation}\label{10} 
\mathcal L = \frac{1}{2} \phi_{,i} \phi^{,i} + \frac{1}{2} \Box \phi \Box
\phi,  
\end{equation}
 where we use the notation $\Box \phi = \phi^{,i}_{\ ,i}$ and $\Delta \phi = 
\phi^{,\mu}_{\ ,\mu}$ ($\mu= 1,2,3$). The field equations are $\Box \Box \phi
- \Box \phi = 0$. We do not discuss the physical relevance of such a 
theory (therefore, coupling constants have been omitted). The inclusion of 
a potential, in particular a mass term, is trivial and does not lead to 
significant modifications of our discussion. In what follows, we 
will refer to the theory based on the above Lagrangian 
as {\it example I}.

The momenta are found from (\ref{5}) in the form 
\begin{equation} \label{11} 
\pi = \dot \phi - \Box \dot \phi, \ \ \ p^m = \delta^m_0 \Box \phi. 
\end{equation}
It turns out to be  convenient to simplify the notations for the time 
components of $p^m$ and $\psi_{m} = \phi_{,m}$, and to use 
$p^0 = p$ and $\psi_0 = \psi$. Thus, we have $\psi = \dot \phi$ (as in 
the Ostrogradski formulation), as well as $\psi_{\mu} = \phi_{,\mu}$. 
Those three relations do not contain velocities, and must therefore 
be considered as constraints. Next, we have $p = \Box \phi = 
\dot \psi + \Delta \phi $ and $\pi = \psi - \dot p $, both containing 
 velocities, as well as $p^{\mu} = 0$, which are constraints again. 

As a result, we have the following 6 constraints 
\begin{equation} \label{12} 
\Phi_{\mu} = \psi_{\mu} - \phi_{,\mu}, \ \ \ \Psi^{\mu} = p^{\mu}, 
\end{equation}
which satisfy 
\begin{equation} \label{13} 
[\Phi_{\mu}, \Psi^{\nu}] = \delta^{\nu}_{\mu} \delta, 
\end{equation}
as well as $[\Phi_{\mu}, \Phi_{\nu}] = [\Psi^{\mu},\Psi^{\nu}] = 0$, 
where we omit the arguments for simplicity of the notation. Thus, we 
are dealing with second class constraints, which have to be dealt with 
by the introduction of the Dirac bracket (see \cite{dirac}). For the 
specific structure (\ref{13}), it is not hard to show that the 
Dirac bracket can be written in the form 
\begin{equation} \label{14}
[A, B]^* = [A,B] + [A,\Phi_{\mu}] [\Psi^{\mu}, B]-
[A,\Psi^{\mu}] [\Phi_{\mu}, B]. 
\end{equation}
Note that apart from a summation over 
$\mu$, there is also an  integration over the argument 
of $\Phi_{\mu}(z)$ and $\Psi^{\mu}(z)$ involved, which is suppressed by 
our simplified notation.  

We now easily find the following Dirac bracket relations 
\begin{equation}\label{15} 
[\phi, A]^* = [\phi,A],\ \ [\psi,A]^* = [\psi,A],\ \ [\psi_{\mu},\pi]^* = 
\delta_{,\mu},\ \ [p,\pi]^* = 0,\ \ [p^{\mu}, \psi_{\nu}]^* = 0,  
\end{equation}
where $A$ is arbitrary. As it turns out, the relations are exactly those 
that could have been assumed right from the start if one would not have 
considered $\psi_{\mu}$ and $p^{\mu}$ as canonical variables.
This is rather a coincidence, however, as we will see in the next example. 
(The notation $\delta_{,\mu}$ is to be interpreted as $f \delta_{,\mu} = 
- f_{,\mu} \delta$ for a function $f$. The omission of the 
arguments is not without danger. Note, for instance, that from 
$[\psi_{\mu}(x),\pi(y)] = \delta_{,\mu}(x-y)$, we find $[\pi(y),
\psi_{\mu}(x)]= - \delta_{,\mu}(x-y) = \delta_{,\mu}(y-x)$, and thus, 
in short notation, $[\pi, \psi_{\mu}] = \delta_{,\mu}$, contrary to 
what could have been expected from the initial relation 
$[\psi_{\mu}, \pi] = \delta_{,\mu}$ and the antisymmetry 
of the Poisson bracket.)

In any case, we can now impose the constraints as strong relations 
between the variables, and thus eliminate $\psi_{\mu}$ and $p^{\mu}$ 
as independent field variables. Then, we use (\ref{8}) in order to 
write down the Hamiltonian. The velocities (of the remaining 
variables $\phi$ and $\psi$) are easily expressed in terms of 
the momenta as $\dot \phi = \psi $ and $\dot \psi = p - \Delta \phi$. 
We find 
\begin{equation} \label{16}
\mathcal H = \int (\pi \psi + \frac{1}{2} p^2 - p \Delta \phi - \frac{1}{2} 
\psi^2 - \frac{1}{2} \phi_{,\mu}\phi^{,\mu}) \de^3 x. 
\end{equation}
With the help of (\ref{15}), we find 
\begin{equation} \label{17}
[\mathcal H, \phi]^* = -\psi = -\dot \phi 
\end{equation}
\begin{equation}\label{18}
[\mathcal H, \pi]^* = - \Delta p + \Delta \phi = 
-( \Box \Box \phi - \Box \phi) - ( \ddot \phi - \Box \ddot \phi) 
= - \dot \pi, 
\end{equation}
\begin{equation} \label{19} 
[\mathcal H, \psi]^* = - p + \Delta \phi = - \dot \psi, \ \ \ 
[\mathcal H, p]^*= \pi - \psi = - \dot p.  
\end{equation}
Thus, the Hamiltonian does indeed generate the time evolution of 
the  fields. To conclude, 
despite the fact that additional pairs of variables $(\psi_{\mu}, \pi^{\mu})$ 
revealed themselves as irrelevant, the formalism has nevertheless 
successfully passed its first test. 

We now start form the Lagrangian 
\begin{equation} \label{20}
\mathcal L = \frac{1}{2} \phi_{,i} \phi^{,i} 
+ \frac{1}{2} \phi_{,i,m}\phi^{,i,m}. 
\end{equation}
This theory,  which will be referred to as {\it example II}, 
is equivalent to the previous one in the sense that 
it leads to the same field equations. It differs,  however, by a 
four divergence and therefore, differences in the Hamiltonian theory 
will arise. We find $\pi = \dot \phi - \Box \dot \phi$, $ p^m= 
\dot \phi^{,m}$, and thus, writing again $\psi = \psi_0$ and $p = p^0$,  
we have $\pi = \psi - p^{\mu}_{\ ,\mu} 
- \dot p$ and  $p = \dot \psi$, which are relations involving 
velocities. In addition, we have the constraints 
\begin{equation} \label{21} 
\Phi_{\mu} = \psi_{\mu} - \phi_{,\mu}, \ \ \ 
\Psi^{\mu} = p^{\mu}- \psi^{,\mu}, 
\end{equation}
which satisfy again $[\Phi_{\mu}, \Phi_{\nu}]= [\Psi^{\mu}, \Psi^{\nu}]$ and 
\begin{equation} \label{22} 
[\Phi_{\mu}, \Psi^{\nu}] = \delta^{\nu}_{\mu} \delta.  
\end{equation}
Although the constraints satisfy the same Poisson structure as in example I, 
they are 
nevertheless fundamentally different because of the explicit 
occurrence of $\psi$. Indeed, we now find the following Dirac brackets 
\begin{equation}\label{23} 
[\phi, A]^* = [\phi,A],\ \ [\psi,A]^* = [\psi,A],\ \ [\psi_{\mu},\pi]^* = 
\delta_{,\mu},\ \ [p,\pi]^* = - \Delta \delta, 
\ \ [p^{\mu}, \psi_{\nu}]^* = 0  
\end{equation}
for arbitrary $A$. They are identical to (\ref{15}), except for the 
relation $[p,\pi]^* = - \Delta \delta $, which is of course again 
a symbolic notation for $ f (\Delta \delta)  = (\Delta f) \delta$. 
Imposing the constraints as strong relations in $\mathcal H$, we find 
\begin{equation}\label{24} 
\mathcal H = \int ( 
\pi \psi + \frac{1}{2} p^2 - \frac{1}{2} \psi^2 - \frac{1}{2} \phi_{,\mu} 
\phi^{,\mu} - \frac{1}{2} \phi_{,\mu,\nu} \phi^{,\mu,\nu} 
)\de^3 x, 
\end{equation}
where the velocities have been eliminated by $\dot \psi = p $ and 
$\dot \psi_{\mu} = \pi_{\mu} = \psi_{,\mu}$, as well as $\dot \phi = \psi$. 
We now easily derive 
\begin{equation} \label{25}
[\mathcal H, \phi]^* = -\psi = -\dot \phi 
\end{equation}
\begin{equation}\label{26}
[\mathcal H, \pi]^* = - \Delta p + \Delta \phi - \Delta \Delta \phi = 
- \Delta \ddot \phi + \Delta \phi + \Delta \Delta \phi  
=  (\Box \phi - \Box \Box \phi) - (\ddot \phi - \Box \ddot \phi)   
= - \dot \pi, 
\end{equation}
\begin{equation} \label{27} 
[\mathcal H, \psi]^* = - p  = - \dot \psi, \ \ \ 
[\mathcal H, p]^*= \Delta \psi + \pi - \psi = - \dot p.  
\end{equation}
Again, the formalism works perfectly well. As before, the additional 
variables $(\psi_{\mu}, p^{\mu})$ could be eliminated after the introduction 
of the Dirac brackets. It is expected that this is a general feature of 
our formalism, if there is any hope for it to be equivalent 
to the Ostrogradski formulation based on only two pairs of variables. 
On the other hand, it should be noted that, in contrast to example I, 
in the present  case, we could not have anticipated the relation 
$[p, \pi]^* = - \Delta \delta$. If we simply ignore the variables 
$\pi^{\mu}$ and $\psi_{\mu}$, then this relation cannot be derived, since 
we need the complete set of constraints (\ref{21}) to get the 
correct Dirac brackets. Finally, in order to avoid confusion, we should 
mention that in the  title of this section, we use the term
\textit{unconstrained} in the sense that in the conventional Ostrogradski 
formulation, those theories are indeed free of constraints. In our 
modified formalism, there will always be at least those constraints that 
eliminate the variables $\psi_{\mu}$ and $p^{\mu}$. In a 
{\it constrained} theory, there will be additional constraints.

\section{Constrained theory} \label{constr}

We now consider the Lagrangian 
\begin{equation} \label{28}
\mathcal L = \frac{1}{2} \phi_{,i} \phi^{,i} + \alpha \phi \Box \phi. 
\end{equation}
This theory ({\it example III}) 
is equivalent to the conventional first order scalar field 
theory and we can thus expect that the application of the second order
formalism leads to a constrained system. The momenta are found in the 
form $\pi = \dot \phi(1-\alpha) = \psi (1-\alpha)$ and $p^m = \delta^m_0
 \alpha \phi$. Thus, apart from the constraints 
\begin{equation} \label{29} 
\Phi_{\mu} = \psi_{\mu} - \phi_{,\mu}, \ \ \ \Psi^{\mu} = p^{\mu}, 
\end{equation}
we now have the additional constraints 
\begin{equation} \label{30}
\Phi = p - \alpha \phi, \ \ \ \Psi = \pi - (1-\alpha) \psi, 
\end{equation}
since none of those relations involves velocities. The constraints are 
all second class. Rather than  deriving directly  the Dirac brackets 
for the system 
of those eight constraints (at each point $x$), it is convenient to 
proceed in two steps. First, we construct the Dirac bracket that eliminates 
the constraints (\ref{29}). (The construction of the  
Dirac brackets by eliminating the constraints in two steps will be 
given at the end of this section.) 

The constraints are identical to those 
of example I, and the result, as we have seen, is simply that we can 
eliminate the variables $\psi_{\mu}$ and $p^{\mu}$ in terms of $\phi_{,\mu}$. 
The Dirac brackets for the remaining variables are identical to the 
initial Poisson bracket,  see (\ref{15}). We will therefore retain the 
notation $[A,B]$.  

In a second step, we turn to the constraints (\ref{30}), which  satisfy 
\begin{equation} \label{31}
[\Phi, \Psi] = (1-2 \alpha) \delta. 
\end{equation}
The corresponding Dirac bracket is easily shown to be of the form 
\begin{equation} \label{32} 
[A,B]^* = [A,B] + \frac{1}{1-2\alpha} [A,\Phi][\Psi,B]
- \frac{1}{1-2\alpha} [A,\Psi][\Phi,B], 
\end{equation}
where again an integration is suppressed by our 
notation. We find the following relations 
\begin{equation} \label{33}
[\pi, \phi]^* = - \frac{1-\alpha}{1-2\alpha}\ \delta,  \ \ \ 
[\phi,p]^* = [\pi, \psi]^* = 0,
\ \ \ [\pi,p]^* = -\alpha \frac{1-\alpha}{1-2 \alpha}\ \delta , \ \  \ 
[\phi,\psi]^* = \frac{1}{1-2 \alpha}\ \delta.
\end{equation}
The Hamiltonian is found from  (\ref{8}) upon  imposing the constraints 
as strong relations. The result is 
\begin{equation} \label{34}
\mathcal H = \int \left( \frac{\frac{1}{2} - \alpha}{(1-\alpha)^2}
\ \pi^2 - (\frac{1}{2} - \alpha) \phi_{,\mu} \phi^{,\mu} - 
\alpha(\phi \phi^{,\mu})_{,\mu}\right)\de^3 x. 
\end{equation}
If we rescale the momentum $\pi$ by introducing $\hat \pi = 
\frac{1-2 \alpha}{1-\alpha}\ \pi$, such that $[\hat \pi, \phi]^*  
= - \delta $, we can alternatively write 
\begin{equation} \label{35}
\mathcal H = \int\left(\frac{1}{2}\  \frac{1}{1-2 \alpha}
\ \hat \pi^2 - (\frac{1}{2} - \alpha) \phi_{,\mu} \phi^{,\mu} - 
\alpha(\phi \phi^{,\mu})_{,\mu}\right)\de^3 x.   
\end{equation}
Apart from the last term, which is a surface term, the Hamiltonian corresponds
 to the conventional first order Hamiltonian derived from $\mathcal L = 
(\frac{1}{2} - \alpha) \phi_{,i}\phi^{,i}$, which is equal to  (\ref{28}) 
up to a four divergence. 

For the rest, the relations 
\begin{equation}\label{36} 
[\mathcal H, \phi]^* = - \dot \phi, \ \ \ [\mathcal H, \pi]^* = - \dot \pi 
\end{equation}
are easily verified. Thus, our formalism works even for such a strongly 
constrained system. 

The Lagrangian (\ref{28}) for the specific value $\alpha = 1$ can be viewed 
as a toy model that mimics in some sense the Lagrangian of General 
Relativity. Namely, the Lagrangian $\sqrt{-g}\ R$ consists of a part 
containing only first derivatives of the metric and a part that contains 
the metric and its  second derivatives (linearly). The second part equals, 
up to a four divergence, the double of the opposite    of the first part, 
similarly as in (\ref{28}) for $\alpha = 1$. It is indeed the scope  
of our  {\it  exercise} to provide a Hamiltonian formalism that can be 
applied to General Relativity in its explicitely covariant form, in contrast 
to the conventional first order method, where a surface term has to be 
omitted, resulting in an effective Lagrangian that is not explicitely 
covariant. 

At first sight, this may look like an unnecessary complication, 
since the number of variables is initially increased only to be reduced 
again at a later stage by imposing the constraints. Nevertheless, it is 
hoped that despite those computational complications, there will be an 
improvement of clarity in particular concerning the physical meaning of 
the constraints of the theory. Indeed, it is well known that the 
primary and secondary first class constraints arising in generally 
covariant theories are directly related to 
 diffeomorphism  invariance. 
It turns out that those constraints 
can be directly inferred by a straightforward analysis of the corresponding 
Noether currents (that is, the stress-energy tensor). The explicit form  
 of the constraints and their action as generators of coordinate
 transformations has been given in \cite{leclerc} for first order theories, 
and similar relations are easily  derived for second order theories, following 
along the same lines. On the other hand, if we work with an effective,
not explicitely covariant Lagrangian, the relation between constraints and 
generators of coordinate transformations is more obscure and explicit 
calculations have to be performed in order to determine the action of the 
constraints on the fields. For instance, simple relations, like the symmetry 
properties of (\ref{3}), are not valid anymore.   
There will be, of course, alternative relations expressing the 
(hidden) coordinate invariance, 
but those  will not emerge directly from Noether's theorem and have to be 
obtained  more or less by guesswork. 

A second, related, issue concerns the occurrence of surface terms in 
the Hamiltonian. In view of the specific asymptotic behavior 
of the metric, e.g., $g_{00} = 1- m/r$ (for asymptotically flat spacetimes), 
surface integrals occurring in  General Relativity do not always vanish, 
in contrast to conventional field theories.  
 In fact, it is not hard to show from 
(\ref{3}) that 
the only field that explicitly contributes to the integrated field 
momentum is the gravitational field. This raises problems when the 
effective, first order Lagrangian is used in General Relativity, since the 
omission of four divergences in the Lagrangian ultimately leads to
 a modification of $\mathcal H$ by surface terms, see, e.g., (\ref{35}).    
Initially, in the context of canonical quantum gravity, 
certain surface terms in the Hamiltonian 
where simply ignored, since they are 
dynamically irrelevant \cite{dirac2}. Later \cite{dewitt}, it was 
recognized that by the omission of those surface terms, we actually 
omit the complete field energy of the system (such that the resulting 
Hamiltonian vanishes weakly) 
and it was argued, based on comparison with the linearized theory, that 
those terms should not be omitted  (except for 
closed spaces).  This was confirmed in \cite{teitel}, where it was shown 
that without those terms, the Hamiltonian formulation of the theory 
is classically inconsistent, because  the variations $\delta \mathcal H/ 
\delta \phi$ and $\delta \mathcal H/ \delta \pi$ 
cannot be properly defined if those terms 
are missing, and thus, we cannot write down the Hamiltonian equations of 
motion. For a discussion of the treatment of surface terms in the variational 
principle of field theory, see also \cite{tapia2} and \cite{tapia3}.  

Obviously, in view of this situation, 
it seems promising to start directly from 
the full Lagrangian $\sqrt{-g}\ R$, instead of omitting first a 
four-divergence, 
and then eventually reintroduce it again (in the form of a three-divergence) 
into the   Hamiltonian in order to   
get a consistent theory. With  the use of the second order Hamiltonian 
formulation, it should be  possible to proceed strictly canonically, 
without 
ever being in the need to omit or add a surface term. The resulting
Hamiltonian can be interpreted directly as field energy and should generate 
the time evolution of the system, provided we are able to deal consistently 
with all the constraints. 
 In asymptotically flat spacetimes, 
it does not vanish weakly. We will outline this procedure in section \ref{gr}.

Since it might not be obvious that the construction of the Dirac 
brackets in two steps (namely, first eliminating the constraints (\ref{29})
and then the constraints (\ref{30})) leads indeed to the same result than the 
construction following directly the procedure of Dirac, we close this 
section by giving a  justification of this procedure for an arbitrary theory.  

Suppose we have second class constraints $\Psi^i,  \Phi^{\alpha}$, where
it is irrelevant whether the labels $i, \alpha$ run over a finite set
(index) or an infinite set (like the argument $x$). Let the Poisson
brackets be given by
\begin{displaymath}
 [ \Psi^i,\Psi^k] = C^{ik}, \ \ [\Phi^{\alpha}, \Phi^{\beta}] = D^{\alpha
\beta},
\end{displaymath}
with invertible, antisymmetric $C^{ik}$, $D^{\alpha\beta}$, the inverse
being denoted by $C_{ik}$ and $D_{\alpha\beta}$ respectively. Nothing
is assumed for $[\Psi^i,\Phi^{\alpha}]$. They  may or may not vanish.

Let $A,B$ be any expression of the canonical variables (fields (or
coordinates) and momenta). Then, in a first step, we define
\begin{displaymath}
[A,B]^* = [A,B] - [A,\Psi^i]C_{ik}[\Psi^k, B],
\end{displaymath}
from which we easily find $[A,\Psi^m]^* = [\Psi^m, B]^* = 0$ for
any of the $\Psi^m$'s and for arbitrary  $A,B$.

In a second step, define
\begin{displaymath}
[A,B]^{**} = [A,B]^* - [A,\Phi^\alpha]^* D_{\alpha\beta}[\Phi^\beta, B]^*.
\end{displaymath}
Obviously, we have $[A, \Phi^{\gamma}]^{**} = [\Phi^{\gamma}, B]^{**} =0$ for
any of the $\Phi^{\alpha}$'s and for arbitrary $A,B$. But since we also
have $[A,\Psi^i]^* = [\Psi^i,B]^* = 0$ for any $A,B$ (and thus,
in particular, e.g., $[\Psi^i, \Phi^{\alpha}]^* = 0$), we find that trivially
$[A,\Psi^i]^{**} = [\Psi^i,B]^{**} = 0$ for any $A,B$.

Moreover, from the above construction, it is also clear that
for those  $A,B$ that  commute  with all of the constraints (i.e.,
$[A,\Psi^i] = [A,\Phi^{\alpha}] = 0$, and similar for $B$), we have
$[A,B]^{**} = [A,B]$.

Summarizing, the bracket $[\ ,\ ]^{**}$ has the following
properties: (1) $[A, \Xi^M]^{**} = 0$ for all of the second class constraints
$\Xi^M = (\Psi^i, \Phi^{\alpha})$ and for arbitrary $A$. (2) If $A,B$
commute with all of the constraints ($[A,\Xi^M] = [B,\Xi^M]=0$), then
$[A,B]^{**} = [A,B]$. But those are exactly the properties that
define the Dirac bracket. 

In other words, the bracket $[\ ,\ ]^{**}$ has exactly the same properties
as the bracket
\begin{displaymath}
[A,B]^{\#} = [A,B] - [A,\Xi^M]E_{MK}[\Xi^K, B],
\end{displaymath}
where $E^{MK} = [\Xi^M, \Xi^K]$, and $E_{MK}$ the inverse of $E^{MK}$.  But
since the bracket defined by the above properties (1) and (2) is unique, we
must have $[A,B]^{\#} = [A,B]^{**}$. There is thus no need to show this
explicitely.

This justifies the two step construction  of
the Dirac brackets.

\section{Original Ostrogradski formulation} \label{ostro}

We started our investigation from the explicit expression of the canonical 
stress-energy tensor. From its structure, it was most natural 
to introduce $\phi$ and $\psi_{m} = \phi_{,m}$ as independent variables and 
to base the canonical Hamiltonian formalism on that. On the other hand, 
the canonical stress-energy tensor is not the only conserved current 
available. As is well known, arbitrary relocalization terms can be added to 
$\tau^k_{\  i}$ without changing the relation $\tau^k_{\ i,k} = 0$. 
In special relativistic theories, not even the integrated momentum 
is  changed by such a procedure. (Gravity, however, provides 
an exception to this, because of the previously mentioned different  
 asymptotic behavior.) In view of those ambiguities concerning in 
particular the energy density, it is  not really  
surprising that there may also exist several Hamiltonians for one and 
the same theory. 

There is a simple way to get at least to two of such Hamiltonian
descriptions. As is well known, in first order theory, the 
momentum  can be derived directly from variation of the action 
functional as $\delta S/ \delta \phi = \partial \mathcal L/ \partial \dot \phi 
= \pi$. Here, the notation  $\delta A$ denotes a  three dimensional 
variation, i.e., if we find for the variation of a functional $A(\phi, \pi)$ 
that $\delta A 
= \int (a_1 \delta \phi  + a_1 \delta \pi) \de^3 x$, then by definition, 
$a_1 = \delta A/ \delta \phi$ and $a_2 = \delta A/ \delta \pi$. 
It turns out that in second order theories, a similar variation of the action 
does not lead to a unique definition of the canonical momenta. 
Indeed, we have 
\begin{equation} \label{37} 
\delta S = \int 
\left(\frac{\partial \mathcal L}{\partial \phi} \delta \phi 
+\frac{\partial \mathcal L}{\partial \phi_{,i}} \delta \phi_{,i} 
+\frac{\partial \mathcal L}{\partial \phi_{,i,k}} \delta \phi_{,i,k}
\right)\de t\ \de^3 x. 
 \end{equation}
Using the field equations $
\partial \mathcal L/ \partial \phi = 
(\partial \mathcal L/ \partial \phi_{,i})_{,i} -  
(\partial \mathcal L/ \partial \phi_{,i,k})_{,i,k}$ in the first term, 
then performing several partial integrations, where three divergences 
can be omitted (surface terms), while the time integration over 
time derivatives can be carried 
out explicitely, one readily finds 
\begin{equation}  \label{38}
\delta S = \int \left( \left[\frac{\partial \mathcal L}{\partial \dot \phi} - 
(\frac{\partial \mathcal L}{\partial \dot \phi_{,i}})_{,i}\right] \delta \phi
+ \frac{\partial \mathcal L}{\partial \dot \phi_{,i}} \delta \phi_{,i} 
\right) \de^3 x, 
\end{equation}
which leads exactly to our previously adapted choice of fields  and momenta 
\begin{eqnarray}\label{39} 
\frac{\delta S}{\delta \phi} &=& 
\frac{\partial \mathcal L}{\partial \dot \phi}-
(\frac{\partial \mathcal L}{\partial \dot \phi_{,i}})_{,i}\ \  \equiv \ \ 
\pi, \\
\frac{\delta S}{\delta \phi_{,i}} &=&\frac{ \partial \mathcal L}{\partial \dot 
\phi_{,i}}\ \  \equiv \ \ p^i. \label{40}
\end{eqnarray}
On the other hand, one can also do with less variables. Writing the integrand 
of the second contribution in (\ref{38}) in the form 
$(\partial \mathcal L/ \partial \dot \phi_{,i} \ \delta \phi)_{,i} - 
(\partial \mathcal L/ \partial \dot \phi_{,i})_{,i} \delta \phi$,
then omitting in the first term of this expression the three divergence 
and carrying out the time differentiation, we find 
\begin{equation} \label{41}
 \delta S = \int \left( \left[
\frac{\partial \mathcal L}{\partial \dot \phi} - 
 (\frac{\partial \mathcal L}{\partial \dot \phi_{,i}})_{,i}
- (\frac{\partial \mathcal L}{\partial \dot \phi_{,\mu}})_{,\mu}
\right] \delta \phi
+ \frac{\partial \mathcal L}{\partial \ddot \phi} 
\delta \dot \phi
\right) \de^3 x, 
\end{equation}
which leads to 
\begin{eqnarray}\label{42} 
\frac{\delta S}{\delta \phi} &=& 
\frac{\partial \mathcal L}{\partial \dot \phi}-
2(\frac{\partial \mathcal L}{\partial \dot \phi_{,\mu}})_{,\mu} 
- (\frac{\partial \mathcal L}{\partial \ddot \phi})_{,0} 
\ \ \equiv\ \  \tilde \pi, \\
\frac{\delta S}{\delta \dot \phi} &=&
\frac{ \partial \mathcal L}{\partial \ddot 
\phi} \ \ \equiv \ \ \tilde p,  \label{43}
\end{eqnarray}
that is, to a theory with variables $(\phi, \psi = \dot \phi)$ and 
corresponding momenta $(\tilde \pi, \tilde p)$. 
Writing down the Hamiltonian for this theory, that is,  
\begin{equation} \label{44} 
\tilde \mathcal  
H = \int (\tilde \pi \dot \phi + \tilde p \dot \psi - \mathcal L)
\de^3 x, 
\end{equation}
and comparing it with our initial Hamiltonian $\mathcal H 
= \int ( \pi \dot \phi + 
p^m \dot \psi_m - \mathcal L) \de^3 x$, we find that the difference  
is given by a surface term 
\begin{equation} \label{45}
\mathcal H = \tilde \mathcal H + \int 
(\frac{\partial \mathcal L}{\partial \dot \phi_{,\mu}} \dot \phi
)_{,\mu} \de^3 x 
= \tilde H + \oint 
\frac{\partial \mathcal L}{\partial \dot \phi_{,\mu}}\dot \phi
\  \de \sigma_{\mu}. 
\end{equation}
Inasfar surface terms are assumed to vanish (as has been assumed 
during the derivation of (\ref{38}) and (\ref{41})), both expressions 
are equal. This does still not mean that the corresponding 
Hamiltonian theories are also equivalent. 

The formulation based on treating the velocities as independent fields 
  is known as 
Ostrogradski formulation,  see, e.g, 
\cite{nesterenko} and \cite{schmidt} and in particular \cite{tapia1}, 
where the formalism has been adapted from 
the case of a finite number of variables to  the case of field theory. 
(The point is 
 that in the finite  case, their is no such thing as a 
spatial derivative, and the distinction between  
the above presented formulations cannot be done  anyway.)

Let us briefly verify the consistency of the formulation based on 
$\tilde \mathcal H$ 
for the three examples we previously  dealt with in 
the alternative formulation. It seems obvious that for the cases of 
the Lagrangian  (\ref{10}) (example I), as well as for the constrained theory 
 (\ref{28}) (example III), both formulations 
 are trivially  equivalent, because of the absence 
of mixed derivatives $\phi_{,\mu,0}$. Indeed, we have for those cases 
$p =  \tilde p$ and $\pi = \tilde \pi$, while the variables 
$\psi_{\mu}$ and $p^{\mu}$ could be eliminated without any changes of 
the Poisson brackets between the remaining variables.

Only the case (\ref{20}) (example II) 
deserves closer examination. We find from 
(\ref{42}) and (\ref{43}) the momenta $\tilde p = \ddot \phi = \dot \psi$ 
and $\tilde \pi = \psi - 2 \Delta \dot \phi - (\ddot \phi)^{\dot{}} = 
\psi - 2 \Delta \psi - \dot {\tilde p}$. The system is thus free 
of constraints. 
The Hamiltonian (\ref{44}) takes the form  
\begin{equation} \label{46} 
\tilde \mathcal H 
=   \int (\tilde \pi \psi + \frac{1}{2} \tilde p^2 - \frac{1}{2}
\psi^2 - \frac{1}{2} \phi_{,\mu}\phi^{,\mu} - \psi_{,\mu}\psi^{,\mu}
- \frac{1}{2} \phi_{,\mu,\nu}\phi^{,\mu,\nu})\de^3 x, 
\end{equation}
where the velocities have been eliminated with $\dot \phi = \psi$ and 
$\dot \psi = \tilde p$. It is needless to say that the only 
non-vanishing fundamental Poisson 
brackets in the present formalism are assumed to be 
\begin{equation} \label{47}
[\tilde \pi, \phi] = - \delta, \ \ \ [\tilde p, \psi] = - \delta.  
\end{equation}
Note, by the way, that if we compare with the momenta 
of the alternative formulation of section \ref{unconstr}, we have 
$\tilde p = p$ and $\tilde \pi = \pi - \Delta \psi$. Therefore, 
from the above Poisson bracket, we can directly \textit{derive}   
 the relation $[p,\pi] = - \Delta \delta$ that arose in 
the other formulation upon  defining the Dirac brackets (see (\ref{23})).  

Next,  from (\ref{46}), we  find 
\begin{equation} \label{48}
[\tilde \mathcal H,\phi] = - \psi = -\dot \phi, \ \ \ \  
[\tilde \mathcal H,\psi] = - \tilde p = - \dot \psi,  
\end{equation}
as well as 
\begin{equation}\label{49}
[\tilde \mathcal H, \tilde \pi ]= \Delta \phi - \Delta \Delta \phi
= - \dot{  \tilde  \pi}, \ \ \ \ 
[\tilde \mathcal H, \tilde p  ]= - \ddot \psi 
= - \dot{ \tilde p},   
\end{equation}
where the field equations have been used in the first relation. 
Thus, the Hamiltonian $\tilde H$ generates indeed the time evolution 
of the phase space variables $\phi,\psi,\tilde \pi$ and $\tilde p$. 

To conclude, we see that both the initial Ostrogradski formulation, 
as well as our modified formulation, lead to  consistent Hamiltonian 
theories  for the simple models analyzed here. Numerically, the 
corresponding Hamiltonians 
differ by a surface integral. 

As outlined at the end of the previous section, we expect certain improvements 
of clarity by the use of our modified formulation. Although the Hamiltonian 
formulation necessarily induces a 3+1 split of spacetime, it seems 
nevertheless in the spirit of a covariant theory to treat $\phi_{,\mu}$ 
and $\dot \phi$ in a symmetric way, at least initially. One advantage of 
such a procedure has already be encountered: The resulting Hamiltonian is 
directly given by the integrated time component of the canonical
stress-energy tensor, i.e., by the time component of the four-momentum 
$\mathcal P_i = \int \tau^k_{\ i} \de \sigma_k$. 
 On the other hand, the Ostrogradski Hamiltonian is 
not a component of anything (it corresponds rather to a generalized Legendre 
transform of the Lagrangian), and its conservation as well as its
identification with the energy have to be established separately. A 
direct relation to the stress-energy tensor, and thus to the Noether current 
corresponding to the translational invariance of the theory, should turn out 
to be profitable in generally covariant theories, where we will have 
to deal with constraints related to diffeomorphism invariance (see
\cite{leclerc}).

\section{General Relativity} \label{gr}

We start from the Lagrangian 
\begin{equation} \label{50}
\mathcal L = \sqrt{-g} R, 
\end{equation}
where for simplicity, we omit the factor $-\frac{1}{2}$ which is necessary 
to get the correct sign for the energy. The field variables, in the 
second order formalism,  are 
$g_{ik}$ and $\psi_{ikm} = g_{ik,m}$. Further, we find from (\ref{5}) 
\begin{eqnarray}\label{51}
\pi^{ik} &=& \frac{\sqrt{-g}}{2} \left[ - \Gamma^i_{lm} g^{lm} g^{k0}   
- \Gamma^k_{lm} g^{lm} g^{i0}   
+ \Gamma^i_{lm} g^{l0} g^{km}   
+ \Gamma^k_{lm} g^{l0} g^{im} \right], \\ \label{52} 
p^{ikm} 
&=& \frac{\sqrt{-g}}{2} 
\left[g^{i0}g^{km} + g^{k0}g^{im} - 2 g^{ik} g^{0m}\right].
  \end{eqnarray}  
The Poisson brackets are assumed to be 
\begin{equation} \label{53}
[g_{ik}, \pi^{lm}] = \delta^{ik}_{lm} \delta, \ \ \ 
[\psi_{ikq}, \pi^{lmr}] = \delta^{ik}_{lm} \delta^r_q \delta, 
\end{equation}
where we use the familiar notation $\delta^{ik}_{lm} = \frac{1}{2}(\delta^i_l
\delta^k_m + \delta^i_m \delta^k_l)$. 

We use again the simplified notation 
$p^{ik} = p^{ik0}$ and $\psi_{ik} = \psi_{ik0} = 
g_{ik,0}$. Obviously, the relations (\ref{51}) and (\ref{52}) are 
all constraints, since the momenta are expressed in terms of $\psi_{ikm}$ 
and $g_{ik}$ (and not in terms of velocities). In addition, we have 
the constraints $\psi_{ik\mu} = g_{ik,\mu}$ (recall that greek indices 
run from $1$ to $3$). As a result, we have a total of 80 constraints, 
and at first sight, most of them seem to be second class. Surprisingly enough, 
with a little bit of patience, the above system can be handled  without 
major difficulties. 

Similar as in example III, we divide the constraints into two groups, 
with the first group consisting of 
\begin{eqnarray} \label{54} 
\Phi_{ik\mu} &=& \psi_{ik\mu} - g_{ik,\mu} \\ \label{55} 
\Psi^{ik\mu} &=& 
p^{ik\mu} - \frac{\sqrt{-g}}{2} \left[ g^{i0}g^{k\mu} 
+ g^{k0}g^{i\mu} - 2 g^{ik} g^{0\mu}\right], 
\end{eqnarray}  
while the second group contains the remaining constraints, 
\begin{eqnarray} 
\Phi^{ik} &=& 
\pi^{ik} - \frac{\sqrt{-g}}{2} \left[ - \Gamma^i_{lm} g^{lm} g^{k0} -  
 \Gamma^k_{lm} g^{lm} g^{i0}   
+ \Gamma^i_{lm} g^{l0} g^{km}   
+ \Gamma^k_{lm} g^{l0} g^{im} \right] \label{56} \\ \label{57} 
\Psi^{ik} &=& p^{ik} - \sqrt{-g}\left[g^{i0}g^{k0} - g^{ik} g^{00}\right].  
\end{eqnarray}
We begin with the first group. Those constraints satisfy 
\begin{equation} \label{58} 
[\Phi_{ik\mu}, \Psi^{lm \nu}] = \delta^{ik}_{lm} \delta^{\nu}_{\mu} \delta,   
\end{equation}
as well as $[\Phi_{ik\mu}, \Phi_{lm \nu}] = [\Psi^{ik\mu}, \Psi^{lm \nu}] = 0$.
This is very similar to our 
previous examples, and the corresponding Dirac bracket reads 
\begin{equation}\label{59}
[A,B]^* = [A,B] + [A,\Phi_{ikm}][\Psi^{ikm}, B] - 
 [A,\Psi^{ikm}][\Phi_{ikm}, B], 
\end{equation}  
as is easily verified. (One has to check that the Dirac brackets between  
any quantity and any of the (above) constraints vanishes.) 
We can now eliminate the variables $\psi_{ik\mu}$ and $p^{ik\mu}$ 
by imposing the constraints as strong relations. 
It is not hard to verify from (\ref{59}) that for the remaining 
variables, we have again that the Dirac bracket is identical 
to the initial Poisson bracket. Indeed, we have $[A,g_{ik}]^* = 
[A,g_{ik}]$,  $[A, \psi_{ik}]^* = [A,\psi_{ik}]$ for any $A$ etc., see 
the corresponding relations (\ref{15}) of section \ref{unconstr}.  
In 
particular, we can also check that $[\pi^{ik}, p^{lm} ]^* = 0$, 
just like in  examples I and III. 

To conclude, the 60 constraints (\ref{54}) and (\ref{55}) can be 
imposed strongly (that is, $\psi_{ik\mu}$ and $p^{ik\mu}$ are 
eliminated), and the remaining brackets remain unchanged. As in example III, 
we will denote them  again  by  $[A ,B ]$, without star.

We are thus left with the  20 constraints (\ref{56}) and  (\ref{57}). 
At this stage, the phase space variables are $g_{ik}, \psi_{ik}, \pi^{ik}$ 
and $p^{ik}$. 
Most constraints turn out to be second class (note that some components of 
$\Gamma^i_{kl}$ depend on $\psi_{ik} = g_{ik,0}$ and have non-vanishing 
brackets with $p^{ik}$). The explicit calculations are quite long, and we 
will only present  partial results here. First, we notice that we have 
\begin{equation} \label{60} 
[\Psi^{ik}, \Psi^{lm} ] = 0.  
\end{equation}
Further, we can derive the relations, 
\begin{equation} \label{61} 
[\Phi^{ik}, \Psi^{0m}] =  [\Psi^{ik}, \Phi^{0m}] = 0.  
\end{equation} 
In particular therefore,  $\Psi^{0i}$ is first class.   
Let us write the remaining brackets in the form 
\begin{equation} \label{62}
[\Phi^{\lambda \delta}, \Psi^{\mu\nu}] = G^{\lambda \delta \mu\nu}\ \delta, 
\ \ \ [\Phi^{\lambda \delta}, \Phi^{\mu\nu}] = 
H^{\lambda \delta \mu\nu}\ \delta.  
\end{equation}
There are further non-vanishing 
 brackets $[\Phi^{\mu\nu}, \Phi^{0i}]$ and $[\Phi^{0i}, \Phi^{0k}]$, 
for which we do not introduce  specific symbols. 
Note, however, that $\Phi^{0i}$ are not  first class. 
It is relatively easy to show that  $G^{\lambda\delta \mu\nu}$ is given by 
the following expression  
\begin{equation} \label{63} 
G^{\lambda\delta \mu\nu} = \sqrt{-g}\left[
\frac{1}{2} g^{\mu\nu} g^{\lambda 0} g^{\delta 0} - \frac{1}{2} 
g^{00} g^{\lambda \delta} g^{\mu\nu} + \frac{1}{2} 
g^{\lambda \mu} g^{\delta \nu} g^{00} + \frac{1}{2} g^{\lambda \delta} 
g^{\mu 0} g^{\nu 0} - g^{\lambda \mu} g^{\delta 0} g^{\nu 0} \right], 
\end{equation}
where the right hand side has still to be symmetrized with respect to 
$\mu \nu$ as well as with respect to $\delta \lambda$. At this point, it is 
convenient to introduce the Arnowitt-Deser-Misner parameterization of 
the metric, that is, we write $g_{00} = 
N^2 - \tilde g_{\mu\nu} N^{\mu} N^{\nu}$, $g_{0\mu} = - N_{\mu}$ and 
$g_{\mu\nu} = - \tilde g_{\mu\nu} $, where $N^{\mu}  = 
\tilde g^{\mu\nu} N_{\mu}$, with  $\tilde g^{\mu\nu}$  defined as 
 inverse of $\tilde g_{\mu\nu}$. We can now write 
\begin{equation} \label{64} 
G^{\lambda\delta \mu\nu} = - \frac{1}{2} N^{-1} 
\sqrt{\tilde g}\left[\tilde g^{\mu\nu} \tilde g^{\lambda \delta} - 
\frac{1}{2} \tilde g^{\lambda \mu} \tilde g^{\delta \nu} 
- \frac{1}{2} \tilde g^{\lambda\nu} \tilde g^{\delta \mu} \right]. 
\end{equation}
Quite interestingly, this is (up to a factor $N^{-1}$) the same {\it metric} 
that appears in the Laplace-Beltrami type term of the Wheeler-DeWitt 
equation (the so-called superspace metric), 
see \cite{dewitt}. Let us also define the {\it inverse} metric 
\begin{equation} \label{65} 
G_{\lambda \delta \mu\nu} = \frac{N}{\sqrt{\tilde g}} \left[
\tilde g_{\delta\mu} \tilde g_{\lambda \nu} 
+ \tilde g_{\delta \nu} \tilde g_{\lambda \mu} - \tilde g_{\delta \lambda}
\tilde g_{\mu\nu}\right], 
\end{equation}
satisfying 
\begin{equation} \label{66} 
G_{\lambda \delta \mu\nu} G^{\mu\nu \alpha \beta} =
\delta^{\alpha\beta}_{\lambda \delta}. 
\end{equation}
As to the other brackets in (\ref{62}), we will not derive them 
explicitely here. In fact,  $H^{\mu\nu\lambda \delta}$ does not 
 simplify a lot. Note that $H^{\mu\nu\lambda \delta}$ 
is antisymmetric with respect to the exchange of the pairs of indices 
$\mu\nu$ and $\lambda \delta$ (in constrast to $G^{\mu\nu\lambda \delta}$,
which is symmetric).  

Having already identified $G_{\mu\nu\lambda\delta}$ as {\it metric}, we 
define $H_{\mu\nu\lambda \delta}$ as 
\begin{equation} \label{66a}
H_{\mu\nu\lambda\delta} = G_{\mu\nu \alpha \beta} H^{\alpha\beta \gamma \rho}
G_{\gamma \rho \lambda \delta}. 
\end{equation}
We introduce the following  Dirac bracket 
\begin{equation} \label{67}
 [A,B]^* =[A,B] + [A,\Phi^{\mu\nu}]\ G_{\mu\nu \lambda \delta}\ 
[\Psi^{\lambda  \delta}, B]
- [A,\Psi^{\mu\nu}]\ G_{\mu\nu \lambda \delta}\ 
[\Phi^{\lambda  \delta}, B] - [A, \Psi^{\mu\nu}]\ H_{\mu\nu\lambda\delta}\  
[\Psi^{\lambda \delta}, B].  
\end{equation}
As always, integration over the arguments of the constraints is understood. 
It is not hard to verify that we have $[\Phi^{\mu\nu}, A]^* 
= [\Psi^{\mu\nu},A]^* = 0$ for any $A$. In particular we now 
have $[\Psi^{\mu\nu}, \Phi^{0i}]^*=0$. We can also verify the 
relation  $[\Phi^{0i},\Phi^{0k}]^* = 0$. As a result, $\Phi^{0i}$ are now 
first class constraints. (This means that the group of constraints (\ref{56}) 
and (\ref{57}) in fact  contained 4 first class constraints, but we did not 
recognize them prior to  the introduction of the Dirac brackets, 
because we did not consider the correct combination of the constraints.) 

All the second class constraints have now been eliminated, and the remaining 
set of variables can be chosen to be $(g_{\mu\nu}, \psi_{\mu\nu}, 
g_{0i}, \pi^{0i}, \psi_{0i}, p^{0i})$. Note that we have chosen
$\psi_{\mu\nu}$ instead of $\pi^{\mu\nu}$, because it is easier to eliminate 
$\pi^{\mu\nu}$ than $\psi_{\mu\nu}$ (see (\ref{56}). This is,  however, merely 
a matter of convenience, and once we have explicitely 
evaluated the Dirac brackets (\ref{67}) 
between all the variables, we can easily reintroduce $\pi^{\mu\nu}$ at any 
stage. We will not  perform this  task completely here, but a few 
relations will be given below. 

The canonical Hamiltonian is constructed from 
\begin{equation} \label{68} 
\mathcal H = \int \left(\pi^{ik} \dot g_{ik} + p^{ikm} \dot \psi_{ikm}
- \mathcal L\right) \de^3 x, 
\end{equation}
which has to be expressed 
in terms of $(g_{\mu\nu}, \psi_{\mu\nu}, g_{0i}, \pi^{0i},\psi_{0i})$. 
Taking into account the first class constraints $\Phi^{0i}$ and $\Psi^{0i}$, 
we find for the total Hamiltonian  
\begin{equation} \label{69} 
\mathcal H_T = \mathcal H + \int \left(\lambda_i\ \Phi^{0i} + 
\mu_i\  \Psi^{0i}\right)  \de^3 x. 
\end{equation}
From this Hamiltonian, properly expressed in terms of the independent 
variables, together with the Dirac brackets (\ref{67}), we can check for 
eventual secondary constraints. Note that $\Psi^{0i} = p^{0i}$. It 
turns out that $p^{0i}$ commutates with the Hamiltonian and does not 
generate secondary constraints. On the other hand,
$\Psi^{0i}$ will generate four  secondary constraints, 
the so-called Hamiltonian 
constraints. The explicit calculations are straightforward, but rather 
lengthy, and will not be carried out here.

Nevertheless, we will give one Dirac bracket explicitely, in order to 
compare with the corresponding result of the conventional first order 
approach. As is easily shown, we have $[\Phi^{\mu\nu}, \psi_{\lambda \delta}] 
= 0$ as well as $[\Psi^{\mu\nu}, g_{\lambda \delta}] = 0$. Further, 
we find $[\Psi^{\mu\nu}, \psi_{\lambda\delta}] = 
- \delta^{\mu\nu}_{\lambda\delta} \delta$ and 
$[\Phi^{\mu\nu}, g_{\lambda\delta}] 
= - \delta^{\mu\nu}_{\lambda\delta} \delta$. From those relations, we can 
 evaluate the Dirac bracket 
\begin{equation} \label{70}
[\psi_{\alpha\beta}, g_{\mu\nu}]^* = G_{\alpha\beta \mu\nu}\ \delta . 
\end{equation}
We recall that the Dirac bracket is the starting point for the 
transition to the second quantized theory (see \cite{dirac}). Therefore  
this relation (which becomes ultimately the commutator 
between $\dot g_{\mu\nu}$ and 
$g_{\mu\nu}$) has to be valid independently of the specific choice of 
variables. Thus, the same relation should hold in the conventional 
first order approach, if only $\dot g_{\mu\nu}$ is expressed in terms of  
the corresponding phase space variables.

Indeed, in the first order approach, we have \cite{dewitt} 
\begin{equation} \label{71} 
\pi_{(1)}^{\mu\nu} = - \sqrt{\tilde g} (K^{\mu\nu} - \tilde g^{\mu\nu} K),  
\end{equation}
where the subscript $(1)$ refers to the choice of variables in the 
first order approach. $K_{\mu\nu}$ is defined as 
\begin{equation} \label{72} 
K_{\mu\nu} = \frac{1}{2} N^{-1} (N_{\mu,\nu} + N_{\nu,\mu} - \dot
{\tilde g}_{\mu\nu}). 
\end{equation}
This can be inverted to 
\begin{equation}\label{73}
\dot {\tilde g}_{\mu\nu} 
= 2 \frac{N}{\sqrt{\tilde g}} \left(\pi_{(1)}^{\alpha\beta} - 
\frac{1}{2} 
\pi_{(1)}^{\gamma\delta}\tilde g_{\gamma\delta}\tilde g^{\alpha\beta}\right)
\tilde g_{\alpha\mu} \tilde g_{\beta\mu} + \dots 
\end{equation}
where the dots indicate that there are additional terms, that do not depend 
on the momenta $\pi^{\alpha\beta}_{(1)}$. The Poisson brackets in the first 
order theory are assumed to be $[\pi^{\alpha\beta}_{(1)}, 
\tilde g_{\mu\nu}]_{(1)} = 
- \delta^{\alpha\beta}_{\mu\nu} \delta$. It is now an easy task to derive 
the symbolic relation 
\begin{equation}\label{74}
[\dot g_{\alpha\beta}, g_{\mu \nu}]_{(1)} 
= G_{\alpha\beta \mu\nu} \delta,  
\end{equation}  
which holds if $\dot g_{\alpha\beta}$ is expressed properly in terms of 
$\pi^{\alpha\beta}_{(1)}$. (Note that $g_{\mu\nu} = \tilde g_{\mu\nu}$ 
in the signature convention of \cite{dewitt}, and $g_{\mu\nu} = - \tilde 
g_{\mu\nu}$ in our convention, but this  
difference is obviously not relevant  for the final relation.) 
We conclude that  both  
approaches ultimately lead to the same commutator between $g_{\mu\nu}$ and 
$\dot g_{\mu\nu}$. This provides strong evidence that the  elimination of 
 the second class constraints has been done consistently.  

\section{Discussion} \label{diss}

In the previous section, we have treated General Relativity as a 
constrained second order field theory. We could successfully eliminate the 
second class constraints and the resulting theory is quite similar 
to the conventional first order approach, containing 8 primary first class 
constraints, 4 of which are trivial and are expected not to lead to 
secondary constraints, while the remaining 4 are expected to lead to 
the so-called Hamiltonian constraints, similar as in the conventional 
approach. 

What we have gained at this point is simply the fact that the Hamiltonian 
equals by construction the canonical field energy, including all eventual 
surface terms. It is important that, in order to achieve this, it is not 
only necessary to use a second order formalism, but rather to use our 
specific, modified formalism, since, as we have outlined in section 
\ref{ostro}, the conventional Ostrogradski Hamiltonian differs by 
a surface term from the canonical energy.  

The fact, however, that  the second order formulation allows us to start from 
the manifestly covariant Lagrangian $\sqrt{-g}\  R$ leads to further 
simplifications. As outlined in \cite{leclerc}, the primary as 
well as the secondary first class constraints 
related to diffeomorphism invariance can be found by inspection of the 
Noether currents. Indeed, as a consequence of Noether's theorem, 
for a generally covariant second order theory,  
four relations can be derived merely from invariance under coordinate 
transformations $x^i \rightarrow x^i + \xi^i$, see \cite{goldberg}. 
They are obtained by a 
successive {\it localization} of the coordinate translations \cite{leclerc}. 
The first, obtained from  
$\xi^i = \epsilon^i = const$, is the conservation of $\tau^k_{\ i}$ 
in the form (\ref{2}), i.e., $\tau^k_{\ i,k} = 0$. The second, obtained from 
 $\xi^i = \epsilon^i_{\ k} x^k$ with constant $\epsilon^i_{\ k}$, 
allows for $\tau^k_{\ i}$ to be written in the form (\ref{3}). The third 
and fourth, from  $\xi^i = \epsilon^i_{\ kl} x^kx^l$ and $\xi^i = 
\epsilon^i_{\ klm} x^k x^l x^m$ respectively, lead to the 
mentioned symmetry properties of the brackets in the expression (\ref{3}).    

Those relations, when integrated over a spacelike hypersurface, lead 
directly to the first class constraints that arise as a result of 
the same symmetries. Let us start with the last one, i.e., 
\begin{displaymath}   
\int\left( \frac{1}{2} \left[\frac{\partial \mathcal L}{\partial \phi_{,m,l}} 
(\sigma \phi)^k_{\ i}\right] \right) \de \sigma_{k}.    
\end{displaymath}
As a result of diffeomorphism invariance, the totally symmetric part in 
$mlk$ of the integrand vanishes (see \cite{leclerc} for details). If we 
choose again $n_k = \delta^0_k$ for the normal vector to the  hypersurface,  
then we must have 
\begin{equation} \label{75} 
 \int\left( \frac{1}{2} \left[\frac{\partial \mathcal L}{\partial \phi_{,0,0}} 
(\sigma \phi)^0_{\ i}\right] \right) \de^3 x = 0,     
\end{equation}
 or simply (omitting the factor $1/2$ and the integration for simplicity) 
\begin{equation} \label{76}
p^{0} (\sigma \phi)^0_{\ i} = 0. 
\end{equation}
(Recall that $p^{0(0)} = p^0$ in our notation, see (\ref{4}) and (\ref{5}).)
In particular, for a symmetric tensor field, we have 
$(\sigma g_{lm})^k_{\ i} = 2 (\delta^k_m g_{li} + \delta^k_l g_{im})$, 
and  we find 
\begin{equation} \label{77}
p^{lm0} (\sigma g_{lm})^0_{\ i} = 4 p^{m0}g_{im} = 0, 
\end{equation}
where we recall our additional convention $p^0 = p$ and thus, for the 
tensor case, $p^{ik0} = p^{ik}$. As expected, this is equivalent to 
the primary first class constraint $\Psi^{0i} = p^{0i} = 0 $, see (\ref{57}).  

Similarly, from the antisymmetry in $kl$ of the first bracket in (\ref{3}), 
we find, integrating over $\de \sigma_k$ and choosing $n_k = \delta_k^0$,  
\begin{equation} 
\int \left[-\frac{\partial \mathcal L}{\partial \phi_{,0,0}} \phi_{,i} 
- \frac{1}{2}
\frac{\partial \mathcal L}{\partial \phi_{,0}} (\sigma \phi)^0_{\ i}
-  \frac{\partial \mathcal L}{\partial \phi_{,m,0}}
[(\sigma \phi)^0_{\ i}]_{,m}\right]\de^3 x = 0
\end{equation}
Written in terms of momenta, we find 
\begin{equation} \label{79}
p \phi_{,i} - \frac{1}{2} (\pi + p^k_{\ ,k}) (\sigma \phi)^0_{\ i} 
+ p_m [(\sigma \phi)^0_{\ i}]_{,m} = 0 
\end{equation}
For the metric theory, this can be written in the form 
\begin{equation} \label{80}
\pi^{0i} =  - p^{0ik}_{\ \ \ ,k} - \frac{1}{2} p^{lm} g_{lm,k} g^{ki} 
- 2 p^{0km} g_{kl,m} g^{il}. 
\end{equation}
If we eliminate  $p^{ikm}$ in terms of the 
dynamical variables (i.e, if we impose the second class constraints
(\ref{54}) and (\ref{55}) as well as $\Phi^{\mu\nu}$ and $\Psi^{\mu\nu}$
from (\ref{56}) and (\ref{57})), we find that the above 
relations are  identical to 
the first class constraints $\Phi^{0i} = 0$ from (\ref{56}).  Note, however, 
that the constraints in the form  (\ref{77}) and (\ref{80}) 
will arise in any covariant second order tensor theory, irrespective of 
the specific Lagrangian and of the eventual presence of additional (second 
class) constraints. 

Thus, we have recovered the primary first class constraints directly 
from the Noether relations obtained in \cite{leclerc} 
from the 
successive \textit{localization} of the coordinate 
translations $x^i \rightarrow x^i + \xi^i(x)$. 

Finally, the secondary first class constraints can be obtained 
from the fact that $\tau^k_{\ i}$ can be expressed both  in the 
canonical form (\ref{2}) and  in the form (\ref{3}). In other words, 
for the canonical field momentum, we can write 
\begin{eqnarray} \label{81}
\mathcal P_i &=& \int \tau^0_{\ i} \de^3 x \nonumber \\
&=&\!\! \int\!\! \left(
\left[-\frac{\partial \mathcal L}{\partial \phi_{,0,l}} \phi_{,i} 
\!- \frac{1}{2}
\frac{\partial \mathcal L}{\partial \phi_{,l}} (\sigma \phi)^0_{\ i}
\!-  \frac{\partial \mathcal L}{\partial \phi_{,m,l}}
[(\sigma \phi)^0_{\ i}]_{,m}\right]_{,l} 
\!\! + \frac{1}{2} \left[\frac{\partial \mathcal L}{\partial \phi_{,m,l}} 
(\sigma \phi)^0_{\ i}\right]_{,m,l} \right) \de^3 x.   
\end{eqnarray}
The expression in the second line  is a surface term, as can be shown 
with the help of the symmetry properties of the brackets under 
the integral  (see \cite{leclerc}). In order to express it in 
terms of the canonical momenta $\pi $ and $p^i$ 
(that is, to eliminate the terms containing $\pi^{(\mu)}$ and 
$p^{i(\mu)}$ appearing in (\ref{81})),  
 it is actually preferable to use first the symmetry properties 
and then integrate (that is, change first the order of the indices 
$kl$ in the first term of (\ref{3}) and then integrate, and similar for 
the second term). This is also necessary to find the generators 
of the coordinate transformations, see \cite{leclerc}. Again, the relation 
(\ref{81}) is valid in any generally covariant second order  theory.   

In any case, we see that the secondary constraints express the fact that 
the canonical field momentum $\mathcal P_i$, and thus in particular the 
Hamiltonian $\mathcal H = \mathcal P_0$, is  equal to a surface term. 
Just as in the conventional first order approach, the Hamiltonian 
vanishes weakly up to a surface term. 
While in the first order approach, this has to be 
checked explicitely, in our approach, it can be anticipated right from 
the start, since we work with an explicitely covariant Lagrangian, 
and can therefore make use of the full power of Noether's theorem. 

As a result, there is no need to explicitely evaluate the 
Hamiltonian in the form (\ref{68}), because it will turn out to 
be weakly equal to the above surface term. 
All we have to do is to properly express 
(\ref{81}) in terms of the dynamical variables, and to construct the 
Hamiltonian which will thus consist of a surface term 
and of the primary and secondary first class constraints derived previously.   

The action of the constraints on the variables and their relation to 
the generators of spacetime translations are easily established following 
along the same lines as in \cite{leclerc}, where the corresponding 
analysis has been carried out for first order theories. 
In contrast to our previous manipulations, i.e., the elimination of 
the second class constraints and the introduction of the Dirac 
brackets, which rely heavily on the Hilbert-Einstein Lagrangian, 
the discussion concerning the first class constraints and 
the generators of translations can be carried out  in a general form 
and the results hold for any generally covariant second order field theory.

\section{Conclusions}

We have presented an alternative Hamiltonian formulation for 
field theories based on Lagrangians that contain second derivatives. 
This formulation differs from the conventional Ostrogradski formalism 
in that all four partial derivatives  of the field are considered as 
independent phase space variables, while in the Ostrogradski method, 
only the time derivative is considered in that way. In our formulation, 
the Hamiltonian is by construction equal to the canonical field energy, 
and will differ, in most cases, from the Ostrogradski Hamiltonian by a 
surface term. It turns out that the additional variables  lead to 
second class constraints and can  easily be eliminated with the help of 
the Dirac bracket. The formalism was applied  successfully to several 
constrained and unconstrained second order scalar field theories and 
its equivalence (up to surface terms) to the Ostrogradski formulation was 
established. Finally, the full power  
of our formulation has been demonstrated 
by applying it to General Relativity. While conventionally, General Relativity 
is treated as first order theory, which leads to difficulties concerning 
certain surface terms that are omitted in the Lagrangian, but have to 
be reinserted into the Hamiltonian for consistency, the second order formalism 
allows us to work directly with the explicitly  covariant Lagrangian. 
This way, we avoid not only  the above problems concerning the surface terms, 
but moreover, the expressions for the primary and secondary first class 
constraints as well as 
their action on the field variables and their relation to the generators 
of coordinate transformations can be directly established  from 
the general structure of the Noether currents.


\begin{thebibliography}{12}

\bibitem{goldberg} J.N. Goldberg, Phys. Rev. {\bf 111}, 315 (1958) 
\bibitem{leclerc} M. Leclerc, gr-qc/0608096 

\bibitem{leclerc2} M. Leclerc, gr-qc/0606033 

\bibitem{dirac} P.A.M. Dirac,
\textit{Lectures on Quantum Mechanics}, 
Yeshiva University, New York 1964 

\bibitem{dirac2} P.A.M. Dirac, Phys. Rev. {\bf 114}, 942 (1959) 

\bibitem{dewitt} B.S.  DeWitt, Phys. Rev. {\bf 160}, 1113 (1967) 

\bibitem{teitel} T. Regge and C. Teitelboim, Annals Phys. {\bf 88}, 286
(1974) 

\bibitem{tapia2} V. Tapia, Nuovo Cim. B {\bf 102}, 123 (1988) 
\bibitem{tapia3} V. Tapia, Phys. Lett.  B {\bf 194}, 408 (1987) 

\bibitem{nesterenko} V.V. Nesterenko, J. Phys. A {\bf 22}, 1673 (1989) 
\bibitem{schmidt} H.-J. Schmidt, Phys. Rev. D {\bf 49}, 6354 (1994)  

\bibitem{tapia1} V. Tapia, Nuovo Cim. B {\bf 101}, 183 (1988) 








\end{thebibliography}
\end{document}